\documentclass{article}
\pdfoutput=1

\usepackage{arxiv}
\usepackage{parskip}
\usepackage[utf8]{inputenc} 
\usepackage[T1]{fontenc}    
\usepackage{hyperref}       
\usepackage{url}            
\usepackage{booktabs}       
\usepackage{amsfonts}       
\usepackage{nicefrac}       
\usepackage{microtype}      
\usepackage{lipsum}		
\usepackage{graphicx}
\usepackage{natbib}
\usepackage{doi}
\usepackage[skip=0.333\baselineskip]{caption}
\usepackage{subcaption}
\usepackage{authblk}

\title{Smart Connected Farms and Networked Farmers to Tackle Climate Challenges Impacting Agricultural Production}

\author[1]{Behzad J. Balabaygloo}
\author[2]{Barituka Bekee}
\author[1]{Samuel W. Blair}
\author[4]{Suzanne Fey}
\author[1]{Fateme Fotouhi}
\author[3]{Ashish Gupta}
\author[3]{Kevin Menke}
\author[3]{Anusha Vangala}
\author[2]{Jorge C. M. Palomares}
\author[4]{Aaron Prestholt}
\author[3]{Vishesh K. Tanwar}
\author[5]{Xu Tao}
\author[4]{Matthew E. Carroll}
\author[3]{Sajal Das}
\author[1]{Gil Depaula}
\author[4]{Peter Kyveryga}
\author[1]{Soumik Sarkar}
\author[2]{Michelle Segovia}
\author[5]{Simone Sylvestri}
\author[2]{Corinne Valdivia}
\author[1]{Asheesh K. Singh\thanks{Corresponding author: Asheesh K. Singh, Department of Agronomy, Iowa State University, Ames Iowa\\ Email: singhak@iastate.edu}}

\affil[1]{Iowa State University}
\affil[2]{University of Missouri}
\affil[3]{Missouri University of Science and Technology}
\affil[4]{Iowa Soybean Association}
\affil[5]{University of Kentucky}

\renewcommand{\shorttitle}{\textit{arXiv} Template}

\hypersetup{
pdftitle={A template for the arxiv style},
pdfsubject={q-bio.NC, q-bio.QM},
pdfauthor={David S.~Hippocampus, Elias D.~Striatum},
pdfkeywords={First keyword, Second keyword, More},
}

\begin{document}
\maketitle

\begin{abstract}
To meet the grand challenges of agricultural production including climate change impacts on crop production, a tight integration of social science, technology and agriculture experts including farmers are needed. There are rapid advances in information and communication technology, precision agriculture and data analytics, which are creating a fertile field for the creation of smart connected farms (SCF) and networked farmers. A network and coordinated farmer network provides unique advantages to farmers to enhance farm production and profitability, while tackling adverse climate events. The aim of this article is to provide a comprehensive overview of the state of the art in SCF including the advances in engineering, computer sciences, data sciences, social sciences and economics including data privacy, sharing and technology adoption.
\end{abstract}

\keywords{Cyber-agricultural systems, IoT, Data Analytics, Precision Ag, Edge Computing, Sensors, Farmer Networks, Technology Adoption}

\section{Introduction}
Agriculture is one of the most important industries that directly or indirectly supports a large section of society. With continual challenges in crop production due to climatic variability, there is a need to consider integrating technical, economic, and social dimensions of research to meet the needs of agriculture. Among the most pressing constraints and challenges of modern  agriculture, climate change-related crop yield decrease is among the most important~\cite{kummu2021climate}. For example, climate change has the potential to reduce agricultural productivity from 3 percent to 16 percent by the year 2080, with developing countries observing more dramatic reductions in productivity ranging from 10 to 25 percent~\cite{mahato2014climate}. Crop diseases and other plant stresses may also become more prevalent as temperatures continue to rise and rainfall becomes more variable \cite{mahato2014climate}. This is ushering in a new paradigm of climate-smart agriculture that requires advances in information and communication technology (ICT) in crop production and agricultural research. This review establishes the concept of smart and connected farms and builds on various technologies and concepts, e.g., the internet of things (IoT), cyber-physical systems (CPS), smart and connected communities, and socio-economic factors, all in the context of empowering farmers through smart and connected farms (SCFs).

\section{Information and Communication Technology in Crop Production}
\label{S:1}
Some of the technologies that make ICT include IoT~\cite{holler2014internet,perkel2017internet}, robotics, big data, artificial intelligence, and cyber-agricultural systems. With the additional support of Cloud services, IoT enables the analysis of large historical data, including soil properties, fertilizer distribution, insect arrival rate, temperature and humidity trend, and so on. While facilitating data collection at every stage of crop production, IoT also paves a path to data-driven services for intelligent farming. IoT mainly refers to the interconnection of sensor-embedded devices/equipment and is envisioned to improve the quality and experience of human living, including in smart agriculture. Potential IoT and wireless sensor networks (WSNs) applications have been reported ~\cite{xue2020smart, patil2016model, sakthipriya2014effective, satyanarayana2013wireless, bhat2021big,sanku2021IoTJ,chen2021Percom}. 

Farmers operate at varying farm size scales, which necessitates context-specific management. To compare scales, traditional farming manages farm operations at the field level, while smart farming allows decision-making at a much smaller scale, i.e., per square meter or plants per unit area. This shift is possible due to Cyber-Agricultural Systems (CAS) that include individualized sensing, modeling, and actuation using machine learning (ML) and coordinated teams of drones/robots that are enabled by autonomy~\cite{noauthor_undated-xy,gao2018novel,sarkar2023cyber}. Automation and control systems provide a higher level of sophistication and precision that can improve the profitability and  sustainability of modern production systems~\cite{lowenberg2020economics}. These technologies support software to help reduce inputs by more accurately targeting within-field areas with variable rate application of fertilizer and pesticides~\cite{babcock1998moving}. Consequently, by targeting these field inputs more accurately, farmers can reduce their chemical footprint on the field. The reduction in inputs saves costs by creating a profit potential for farmers. For example, the use of precision agriculture technologies allows farmers to operate in a more sustainable manner and increase their profitability~\cite{bongiovanni2004precision}.  While these examples are focused on the field level, similar approaches and technologies apply to livestock management~\cite{astill2020smart} and controlled environment farms~\cite{guo2015design}. 

Agricultural equipment manufacturers and service providers are developing technologies that implement precision farming using GPS location, application rate monitoring, satellite image analysis, and predictive models for weather and crop health. While individual farmers are increasingly adopting advanced sensing and intelligent equipment (e.g., smart tractors, seed, and chemical sprayers, aerial surveillance), there is still a lack of efficient communications and computing technology that are effectively connecting farmers within a community, thereby preventing them from sharing data, analyses, and practices optimally. Overcoming these challenges will enable better decision-making at farms and increased crop production while collectively improving the overall well-being and quality of farmers' life.
Although farmers and their technology providers collect farm-level spatial data on planting and harvest, community-wide data collection and analysis will be helpful to formulate management responses to production threats that transcend farm boundaries, for example, weed seed dispersal, pesticide dispersion in air, insect-pest migration from farm to farm, disease spore movement through air and water, and soybean cyst nematode~\cite{boyd2009impact, zivan2017primary}. With rapid networking, farm-level information and production threats can be communicated quickly across a wider agricultural community. Farmer communities can often formulate more effective responses to changing climate events with information sharing and decision-making, whereas isolated farmers or those who do not trust their sources of information often delay taking action against  production threats~\cite{obunyali2019farm}.  This necessitates a technology-driven, community-enabled tool for row crop farming that provides community-wide mitigation efforts against such threats.
\section{Farmer Networks, and Smart and Connected Farms}
\label{S:2}
To improve community-wide data sharing between farmers, novel socio-technical platforms are needed to create an SCF network, which also provides the benefit of an early warning system for damaging pests and other crop stressors. In Africa and parts of western Asia, a large network of farmers uses various sensors and networking tools to help predict and map the spread of locusts that cause major damage to affected countries~\cite{cressman2013role}. This network of technology tools and farmer observations aims to predict when plague-level locust populations may arise and to take preemptive action to stop the locusts before they can cause damage that could cause problems with food insecurity~\cite{cressman2013role}. Networked farms, data aggregation, and sharing data across farms and fields are suitable for limiting the effect of harmful pests in networked farms across the world. 

Although community-based SCF are not yet common, farmers have conglomerated together for on-farm testing networks where experiments are done on the lands of participating farmers for the mutual benefit of all farmers. The data from these tests are more relevant and with a higher confidence level than can be tested by farmers individually. These community-based data collection and downstream decision-making have been done through on-farm networks but without IoT, smart sensors, ML, and other communication technologies. Collecting field experiment data from a community of multiple connected farms can be more useful than collecting single farm-level high-precision sensor data because with on-farm experiments, we can draw cause-and-effect relationships about important crop traits and develop a generalized decision framework for subsequent years at broader geographical regions and growing conditions. However, creating technological tools and solutions needs to ensure that they are important for decision-making rather than being a data set alone. This holds true for drone-based imagery, which may not be of much value unless farmers can take an ``action" based on the information they receive from image-based phenotyping. 

\begin{figure}[h]
\centering
\includegraphics[width=0.8\textwidth]{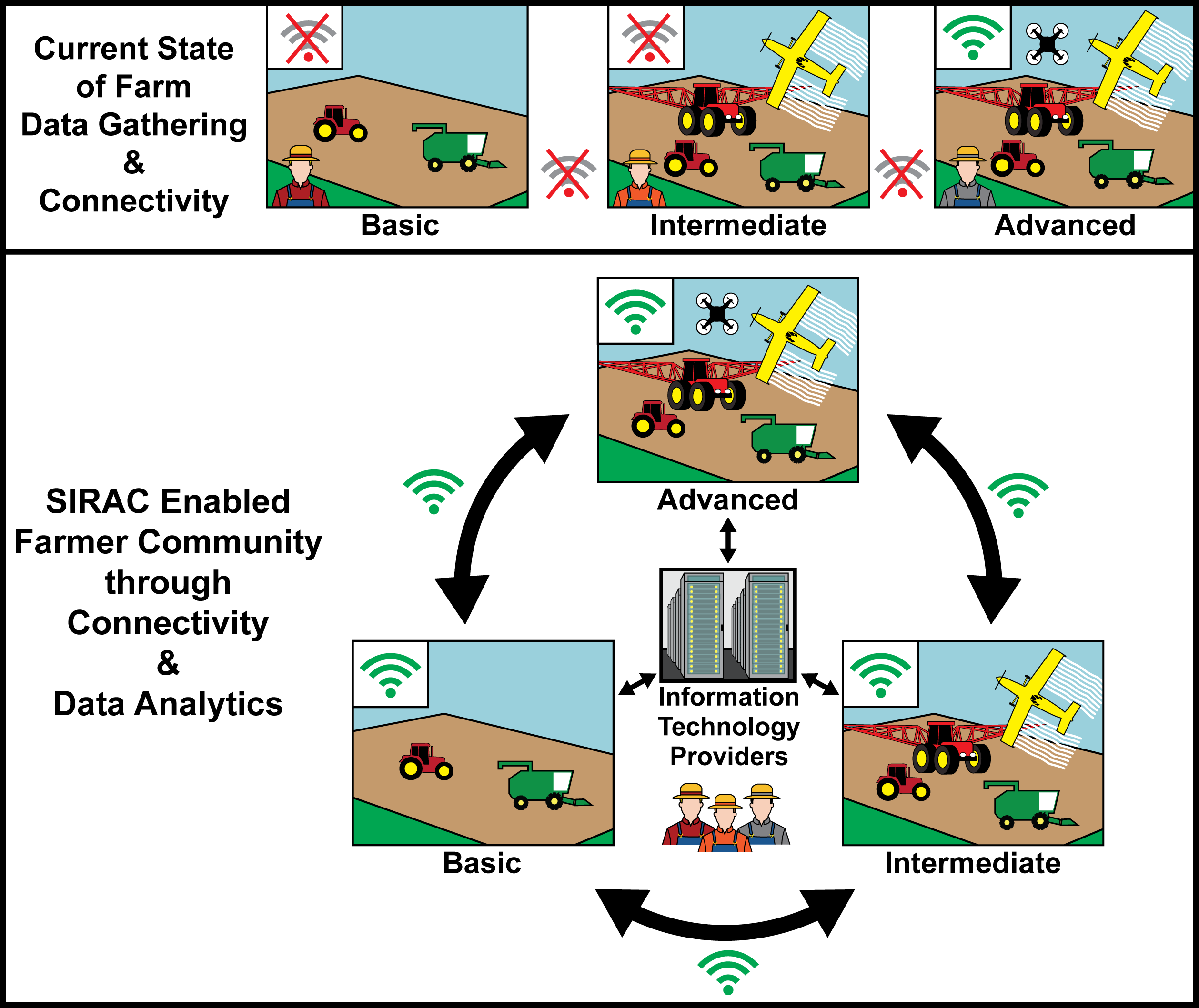}
\caption{\small{\textit{{Example of a smart and connected farm network: Smart Integrated Farm Network for Rural Agricultural Communities (SIRAC) that is designed to tackle complex challenges to agricultural production including climate change.}}}}
\label{SIRAC_central}
\end{figure}

Efforts have begun to create SCF, for example, Smart Integrated Farm Network for Rural Agricultural Communities (SIRAC), that aims to facilitate more effective data sharing, knowledge exchange, and coordinated responses to production threats~\cite{noauthor_undated-xg}. Figure~\ref{SIRAC_central} provides an overview of the SIRAC vision of connecting farms and gathering and sharing information across regions and sensing capabilities. 

Such SCFs will enable real-time monitoring of threats to farm production at a landscape level, which is critical for dealing with pests, diseases, weather, and management issues. SCFs will build on currently successful on-farm research trials that ensure the research and its outcomes are adapted to local farming environments and cultural practices~\cite{kyveryga2019farm}. Technologies designed to create a network of farms and share  information among farmers offer potentially numerous solutions to productivity and sustainability in a changing world~\cite{weiss2020remote}. Furthermore, having a network of smart farms that can share data with one another and utilize remote sensing technologies provides the potential for discovering the extent of climate change’s effect on a farmer’s crop and the viability of growing that crop in areas that have historically supported those crops, potentially meaning a change in land use~\cite{weiss2020remote}. Another networked solution farms offer the balance they can provide between profitability, productivity, and sustainability with aggregated data to better use crop inputs.

\subsection{High-Speed Connections for SCFs}
Only 26 percent of  rural households have broadband connectivity vs. 93.5 percent overall Americans have high-speed internet~\cite{Lopiccalo2021OEA}. Available options for rural households are often limited or more expensive than available in urban areas when considering Dollars/Megabits per second, although newer satellite-based internet options are now available. Agricultural fields have lower coverage because fixed wireless is unavailable, and mobile wireless coverage is available only near highways or populated areas.

Several programs administered by the Federal Communications Commission (FCC) target rural areas to increase the availability of fixed and mobile broadband services for healthcare, schools, and farming. Four such programs to increase the availability of voice, fixed, and mobile broadband services in under-served and rural areas include: Connect America Fund (CAF) for rural areas; Lifeline for low-income consumers; E-rate for schools and libraries, and; the Rural Health Care Program~\cite{noauthor_2012-yk,noauthor_2012-bf,noauthor_2012-ip,noauthor_2010-ea}. The FCC initiatives have identified high-speed connections with at least 25 Mbps downlink and 3 Mbps uplink. According to the FCC, connections with high-speed throughput are considered adequate to upload and download data used in precision farming and field inspection/monitoring tools using Unoccupied Aerial Vehicles (UAVs) or drones and remote sensors~\cite{herr2023unoccupied}. The downlink-to-uplink ratio of 8 to 1 assumes a standard Internet browser information exchange model. 

The programs for increasing connectivity with high-speed connections target family homes, offices, schools, and health centers but not agricultural fields, so other approaches are needed to extend coverage to these remote areas. As of December 2020, the FCC has awarded \$9.2 billion to Internet Service Providers (ISPs) to deploy high-speed internet to unserved homes and businesses in rural areas, while \$6+ billion is still to be awarded for reaching rates of 100Mbps downlink and 20 Mbps uplink~\cite{fccbbawardkcm}.
Extending the high-speed connection to all the fields owned or rented by the operation requires additional networking. Providing an internet access point connection to each field fragment provides a considerable networking challenge. Many farmers manage fields that are variable in size and are dispersed over large areas. Most farming operations include personally owned fields as well as leased fields. Providing ubiquitous coverage of each field is yet another challenge once an internet access point can be provided to each geographically disconnected field. 

\begin{figure}[h]
\centering
\includegraphics[width=0.65\textwidth]{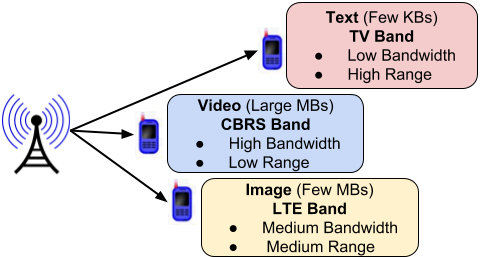}
\caption{\small{\textit{\textbf{{A motivating example of d-DSA paradigm: different bands are preferable depending on the traffic characteristics. }}}}}
\label{dDSA_example}
\end{figure}

Solutions are available for connecting fields but require investment to install new internet services or expand existing ones. Figure~\ref{dDSA_example} shows various methods that can be used to obtain an ISP connection. These include wired alternatives like a Digital Subscriber Line or Fiber to the Home. Wireless alternatives include mobile  wireless provider service over LTE or 5G, Private fixed wireless provider using unlicensed RF bands, or Satellite service. The figure also illustrates connecting field sensors using WiFi technology to create a Local Area Network with internet access that allows sensor connectivity and supports Mobile phone and PC internet service. One common model for accessing precision farming information from agricultural equipment for fields with no internet requires the user to first offload the data from the equipment onto a tablet or PC, and then when they can reach a location with broadband connectivity, the PC connects to a cloud service and uploads the information and then processes and provides analysis and feedback. This model works very well for each equipment manufacturer but has drawbacks for the farmers. Different companies' software does not always integrate well with each other, requiring farmers to use the same manufacturer for all operations or lose the benefits of the data collected with each system and complete farm data integration.

\subsection{Sensors for data collection in SCFs}

SCFs depend on timely and precise sensing (using sensors) of the crop and environment to make informed decisions, for example, when and where to irrigate or spray pesticides. Broadly, the sensors fall into two categories: passive and active. Passive sensors acquire data through light, radiation, heat, or vibrations generated by the physical objects (e.g., crops), while active sensors sense the objects by emitting a signal from their own radiation source and measuring the strength of the reflected or refracted signal~\cite{barmeier2017high}. 

These sensors can be deployed on a wide variety of scales, but are most commonly used on the ground, typically, this would be done by a handheld device at the plant level or field level using a UAV. Further, at the farm and county level, a satellite imaging system is commonly employed to monitor the crops. Depending on farmers' interest in the crop production problem, the resolution, and type of sensor are decided, and appropriate phenotyping agents (robots, drones, satellites) are chosen to allow an increase in throughput as well as area covered~\cite{guo2021uas}. The most commonly used remote sensing modality is a red, green, blue (RGB) digital camera. This is the most easily interpretable data as it senses what the human eye can see. It is also the least expensive sensor when it is being mounted on a UAV. The ease of use both for data collection and downstream processing has made this type of sensor widely used in plant science, and its uses have ranged from biotic~\cite{tetila2017identification,rairdin2022deep} and abiotic~\cite{zhang2017computer,dobbels2019soybean,naik2017real} stress detection, weed detection~\cite{lottes2017uav}, maturity estimation~\cite{trevisan2020high}, and stand counts~\cite{barreto2021automatic}. 

The other commonly used sensor is a multispectral camera, which typically has anywhere from 3-10 bands and usually has bands in the RGB spectrum as well as the red-edge and near-infrared bands. These additional bands can provide additional information about crop health and development and have been used for many use cases listed above~\cite{lottes2017uav} as well as yield prediction~\cite{xu2021cotton}. The increased bands do typically increase the cost of these sensors, but they are still moderately priced and are easily integrated into a UAV. The additional bands can add complexity to downstream analysis and increase the data set size, but many farmers would be accustomed to planting health maps, which are easily generated from this data that can also be acquired via satellites.  

Hyperspectral camera is more sophisticated and typically has 100 or more bands that range from the RGB spectrum to the near-infrared spectrum, but the difference is that there is a much higher spectral resolution with each band typically only being a few nanometers wide. These sensors have been used on UAVs but have had less use than multispectral cameras due to the high cost as well as the complexity of analyzing the data sets. Hyperspectral images have been used in conjunction with machine learning techniques to detect plant stress early in soybeans~\cite{nagasubramanian2018hyperspectral}, as well as for vegetation monitoring in barley~\cite{aasen2015generating}. 

Finally, thermal imaging is another type of passive sensing that uses wavelengths in the far infrared spectrum, typically in the 7,500-14,000 nm wavelengths. Thermal imaging has been used on UAVs, and these sensors account for a moderate cost to purchase but are highly affected by environmental variables. Thermal cameras have been used for disease detection in soybean~\cite{hatton2018remote}, and have been shown to have correlations with biomass and seed yield in drybean~\cite{sankaran2019unmanned}, and also shown for irrigation scheduling in almonds~\cite{garcia2018thermal}.  

Light Detection and Ranging(LiDAR), an example of the active sensor, emits laser pulses that it then uses to detect its distance from an object. These pulses generate three-dimensional point clouds that have been used for terrain mapping from satellites that are utilized to map soil erosion~\cite{gelder2018daily}, and plant biomass estimation from UAVs~\cite{shendryk2020fine}. Processing LiDAR data requires different analysis pipelines than the other sensors because it is no longer working in a 2-dimensional space. 

Different types of sensors will have different time sensitivity. Sensor types with very high time sensitivity, such as minutes or hours, may need a persistent connection to reach their full potential. Sensors with a low time sensitivity, measured in days, may do just fine with periodic connections. Some sensors, such as moisture sensors in the soil, may seem to have a fairly high time sensitivity. However, if the data can be processed locally, instead of in the cloud, this time sensitivity can be mitigated, with the decisions being automated and made in the field.


\subsection{Edge, Fog and Cloud Computing in SCFs}
Despite the advantages of IoT, it is hard to communicate a large volume of sensory (time series or image) data from the agriculture field to the operators or back-end servers (on the Cloud). Such communications not only consume considerable energy but also incur significant communication delays and generate substantial network traffic. A viable solution is {\em edge computing}, where the local processing and storage are available close to the end devices or users~\cite{Li2019, Sirojan2018, esfandiari2021distributed}. Due to the local processing of the tasks near the devices/users, edge computing reduces communication delay and energy consumption for transmitting the collected data. Recent advances in compression-decompression techniques using ML and deep learning (DL) models in Edge Computing can also help reduce the size of the data at the edge devices~\cite{Wang2020, Zhang2017, jiang2017collaborative, esfandiari2021cross}. The edge devices (e.g., sensors deployed in the farmland) have limited resources and can not efficiently execute DL on them. Recently, for large-scale mobile edge computing, an efficient online computation offloading approach via deep reinforcement learning is proposed in~\cite{Hu2022}.

\begin{figure}[h]
\centering
\includegraphics[height=8cm, width=11cm]{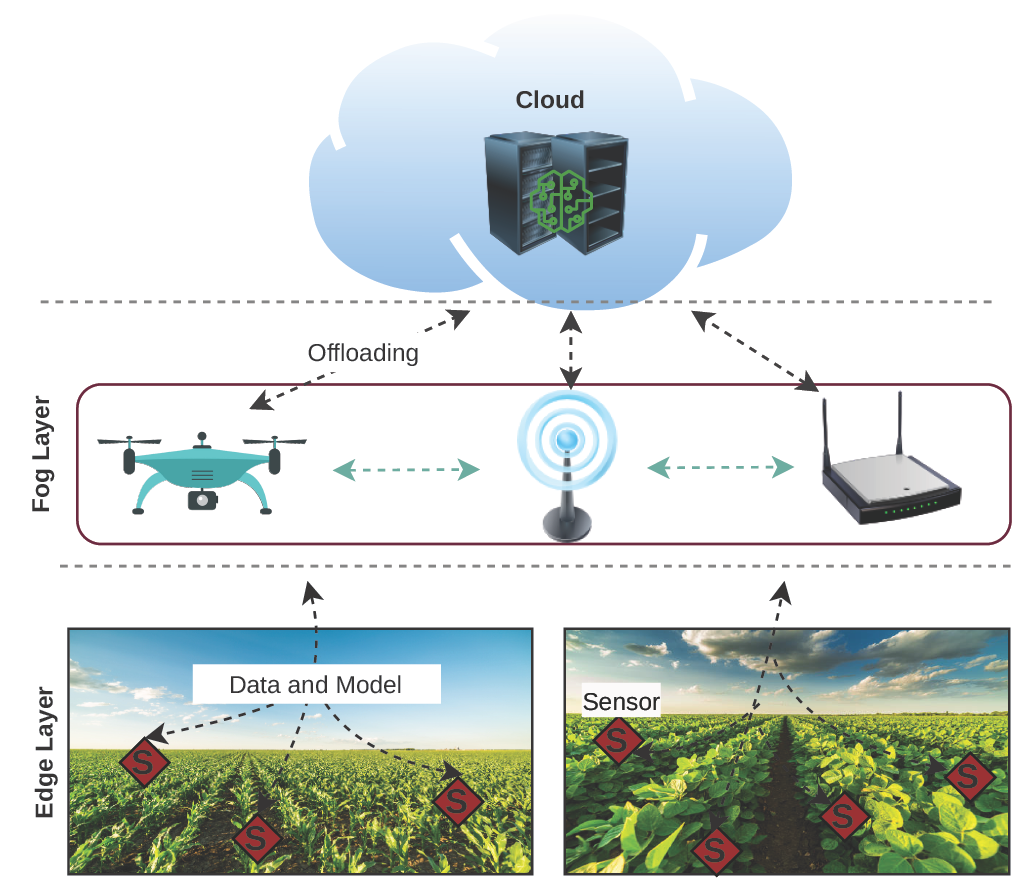}
\caption{An overview of edge, fog, and cloud computing for smart connected farms.}
\label{overview}
\end{figure}

To address the computing limitation of edge devices, an edge-fog architecture is introduced to process IoT data in an efficient way while avoiding the communication delay that occurs in cloud computing~\cite{saha2020dlsense}. Interested readers may refer~\cite{Bellavista2019} for a state-of-the-art survey on fog computing for IoT. Fog devices (e.g., laptops or low-end computing machines) are usually equipped with more resources than edge devices and are mostly located at a lesser communication distance than the cloud. It means that DL with light configuration (few layers and neurons) may be executed efficiently in fog. However, it is not a replacement for the cloud and thus cloud services are still preferable to perform heavy computational tasks in IoT-based farming, e.g., big data (collected from a large number of sensors deployed in the agricultural field) analysis and execution of resource-intensive DL models (e.g., ResNet~\cite{he2016deep}) 
for real-time crop monitoring. Therefore, harnessing the computing capabilities of edge, fog, and cloud through a combined architecture, as depicted in Figure~\ref{overview}, would be the mindful solution to SCFs and smart agriculture. 

\textbf{Case example 1}: Let us consider a scenario with this architecture where several sensors are deployed across many acres of farmland to gather real-time data on soil moisture, crop infection severity, temperature, humidity, etc. Each sensor is attached to a controller (e.g., Arduino or Raspberry Pi) to process the data and it acts as an edge device that transmits this data to a nearby station (e.g., low-end machine) acting as a fog device collecting data from multiple edge devices over the wireless medium. Since the fog devices are capable enough to run light weighted ML and DL models, the farmers may be notified immediately to take necessary actions if any alerting situation is occurred or is predicted. However, for better prediction, historical and genetic information needs to be exploited by the models\cite{shook2021crop,de2017leveraging,shook2021patriot}, which are usually stored in the cloud. The fog devices also offload the data to the cloud in order to receive highly accurate and precise results on insect population, yield prediction, the need for pesticide spray, and so on. The three layers (edge, fog, and cloud) architecture in smart agriculture reduces energy consumption, network traffic, and communication delay~\cite{alharbi2021energy}.

\textbf{Case example 2}: Another way to efficiently collect data from agriculture fields is via participatory sensing or mobile crowdsensing in which the farmers act as human sensors and supply useful information with the help of smartphones. See~\cite{Roy2021, Luo2019} for an energy-aware fog-based framework for data forwarding in mobile crowdsensing; and how to improve IoT data quality in mobile crowdsensing, respectively. Researchers have proposed an energy-efficient data forwarding scheme in fog-based systems with deadline constraints~\cite{Saraswat2020}. Deep neural networks and compression techniques to the edge devices for plant disease detection have been applied~\cite{DeVita2020, DeVita2022}. A drone-based approach to efficiently scouting bugs in orchards using multi-functional nets has been proposed~\cite{Sorbelli2022a, Sorbelli2022b}.

\subsection{Innovative Wireless Technologies for SCFs}

A few recent studies aim to deliver limited Internet connectivity to rural areas. These include DakNet~\cite{daknet-04} and JaldiMAC~\cite{jaldimac,vbts-10,lifenet-11, saha-15}. 
These solutions utilize either short-range technologies (e.g., WiFi, Bluetooth, ZigBee, and 6LowPAN) or long-range solutions (e.g., WiMax, GSM, 3G/4G, LTE/LTE-A, wireless mesh) or a combination thereof. Short-range technologies are generally not adequate for reliable rural connectivity covering wide geographical areas. Conversely, although cellular technologies provide larger transmission coverage and offer promising solutions in the rural context, limited business cases prohibit significant industry investments despite government subsidies~\cite{hasan-14}. 

The recently proposed Long Range Wide Area Network (LoRaWAN) with its scalable star of stars network architecture and simple medium access mechanism, fulfills some of the requirements of providing connectivity in agricultural settings, \textit{i.e.}, long-range communication with low energy consumption~\cite{Shanmuga2020}. The LoRaWAN architecture consists of LoRa Nodes (LNs), LoRa Gateway (LG), and Network Server (NS). Each LG can be connected with a limited number of LNs on a given Spreading Factor (SF) through unlicensed channels. The SFs consume unequal energy and support uneven data transmission rate, and communication range~\cite{Kumari2022}. Thus, the selection of appropriate SF for communicating the time series data from LN to LG helps reduce energy consumption and delay in smart agriculture applications. Although LoRaWAN provides promising applications in the agricultural domain, its very limited data rates make it unfit for some agricultural applications that require large data volumes.  As an example, transferring large hyperspectral images collected by a drone would be infeasible with a low data rate technology such as LoRaWAN~\cite{shah2017designing, shah2018designing}.

To overcome the limitations of existing wireless technologies and better utilize the under-utilized licensed spectrum resources, {\em Dynamic Spectrum Access} (DSA)~\cite{akyildiz-06, song2012dynamic} has emerged as an enabling technology. DSA networking is allowed (or in memoranda) by the United States Federal Communications Commission (FCC) in licensed channels such as TV band~\cite{tvband-10,khalil-17,bahl2009white}, GSM band~\cite{hasan-14}, LTE band~\cite{lte-16}, and CBRS Band~\cite{cbrs}. Similar policies are being adopted in other countries, such as Canada~\cite{canada}, Singapore~\cite{singapore}, South Africa~\cite{southafrica}, UK~\cite{holland2015white}, Malaysia~\cite{malaysia}, Kenya~\cite{kenya}, Namibia~\cite{namibia-1, namibia-2}, and Argentina~\cite{enacom}. DSA technologies have been investigated for rural connectivity, and most are based on TV whitespaces (TVWS)~\cite{bahl2009white, khalil-17, kumar2016toward, liang-08}, GSM whitespaces~\cite{hasan-14} and LTE whitespaces~\cite{surampudi11}. Although existing DSA architectures allow secondary devices to opportunistically access an unoccupied channel, they are restricted to an individual primary band only. Recent studies have shown that  this is neither efficient nor effective under heterogeneous traffic demands and suffer from under- or over-provisioning of spectrum~\cite{shah2017designing, shah2018designing}.

Consider an example in Figure~\ref{dDSA_example}. If the objective is to provide low end-to-end delay, for transmitting a certain text (a few Kilobytes) at large distances (several KMs), a TV band is possibly optimal because it offers high transmission coverage (several KMs) and adequate bandwidth (6 MHz). Whereas, for transmitting a large-sized video at very short distances (a few hundred meters), it may be more efficient to communicate over the CBRS band, which offers very high channel bandwidth (40 MHz) and low (yet sufficiently large) transmission coverage. Similarly, owing to a decent coverage and channel bandwidth, an LTE band is more promising for transmitting a medium-sized image at relatively small distances (a few KMs). 

\begin{figure}[h]
\centering
\includegraphics[width=0.8\textwidth]{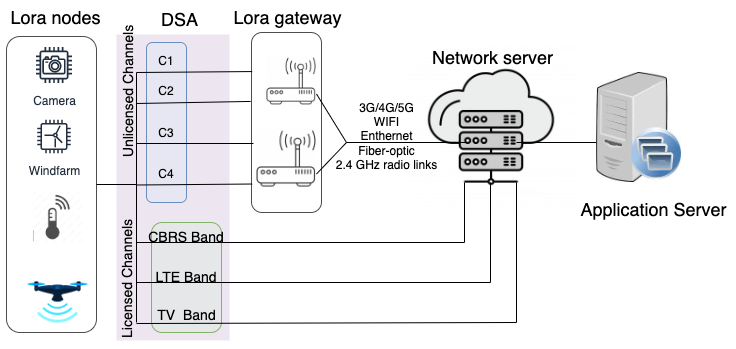}
\caption{\small{\textit{{DSA networking architecture in a farmer network.}}}}
\label{DSA}
\end{figure}

To address the heterogeneous traffic needs, we consider applying DSA on both unlicensed channels used in LoraWAN and licensed primary channels commonly used in the industry as shown in Figure~\ref{DSA}. This approach can improve the efficiency and flexibility of channel selection according to generated data volume, transmission time requirements, and available bands. If the volume of generated data by LNs is small, offloading data to the network server can be performed with unlicensed channels through LoraWAN. Instead, large volume data such as images and videos will be transferred through the free licensed channels to the network server directly.  

\subsection{Data analysis and privacy}
Data analytics is critical in precision agriculture, allowing farmers to adopt data-driven solutions. With numerous benefits, including better yield, reduced waste, and a greater yet precise understanding of environmental factors, data analytics reveals limitless future opportunities in agriculture. Profitable decisions in agriculture could be made when appropriate data are collected and  analyzed timely, key stresses are identified, and prescriptive management actions are efficiently executed~\citep{nagasubramanian2019plant,akintayo2018deep,naik2017real,kar2021self,rairdindeep, singh2021challenges,chiranjeevi2021exploring,naik2017real,ghosal2018explainable,chiozza2021comparative,krause2022using}. These days data has become one of the key elements of smart farms and helps farmers in decision-making and maximizing productivity. Farms can create a lot of data streams, creating a "big data" situation. Big data has 10 dimensions (V's): volume, velocity, variety, veracity, value, variability, vagueness, validity, venue, and vocabulary~\citep{bhat2021big, manyika2011big,kunisch2016big,kamilaris2017review}. 
~\citet{wolfert2017big} and ~\citet{bhat2021big} in their studies comprehensively reviewed the application of big data in agriculture. Public and private companies are building solutions on this platform intending to solve crop production challenges faster, more accurately, and at a larger scale ~\citep{parmley2019development,parmley2019machine,ghosal2019weakly}. Recent advances in data analysis methods such as computer vision, ML, and DL have empowered both researchers and farmers~\citep{singh2018deep,mahmud2021systematic,singh2016machine,riera2021deep}. There are continual improvements in ML methods and their applications, for example, self‐supervised learning has shown improvement in the classification of agriculturally important insect pests in plants~\cite{kar2023self}. For yield prediction, genotypic-topological graph neural network framework built on GraphSAGE has shown promise~\cite{gupta2023agri}.
We have two broad categories of SCF-generated data: Raw sensor data and processed/refined data. The first pass of data analysis will turn the raw data into processed data and will be unique for each type of sensor. This is most likely to use some kind of ML approach. If the raw data is preserved, alternative data analysis methods can be used later, and results can be compared. For the refined data, data analysis will vary by the type of data and could be a simple statistical analysis that could use more complex ML methods for predictions or even a combination of the two.

Data privacy is an important consideration, particularly in the context of SCF, which heavily depends on analyzing and extracting higher-value information from users' raw data. It will be very important to research all local legal obligations about managing users' private data and maintain a privacy agreement with all users to clearly show how all data is used and/or shared. However, the main focus is to invest in techniques that entirely eliminate sharing raw data containing the exact value of data, statistical features, membership, and certain properties ~\citep{liu2021machine}. Recently, distributed Machine Learning (DML), in particular, has had paradigm-shifting impacts on preserving data privacy. In DML, different parties have their private raw data and train a global model by transferring and aggregating the metadata (model parameters or gradients). Therefore, DML can be considered a privacy preserving technique when only metadata is shareable ~\citep{antwi2021privacy}. Two general categories are defined for DML based on the architecture of ML systems; centralized distributed learning and decentralized distributed learning. Transferring the higher-value information (metadata) in Federated Learning (FL) ~\citep{konevcny2016federated, bonawitz2019towards, nguyen2021federated}, the most famous method in centralized distributed learning, happens between the parties and a parameter server (cloud). Then, the server aggregates the metadata to update the global model ~\citep{li2021survey}. There are two different settings for federated learning; cross-device, which involves a large number of IoT devices, and cross-silo containing a small number of reliable clients ~\citep{kairouz2021advances}. Due to the inherent feature of agriculture data such as weather data, soil data, and crop management data, which are scattered and siloed in different servers,~\citet{manoj2022federated} applied a federated averaging technique to train a ML model for soybean yield prediction. Similarly, using the federated learning-based method, researchers could detect the intrusion securely in smart agriculture ~\citep{friha2022felids}.

\begin{figure}[t]
\centering\includegraphics[width=1\linewidth]{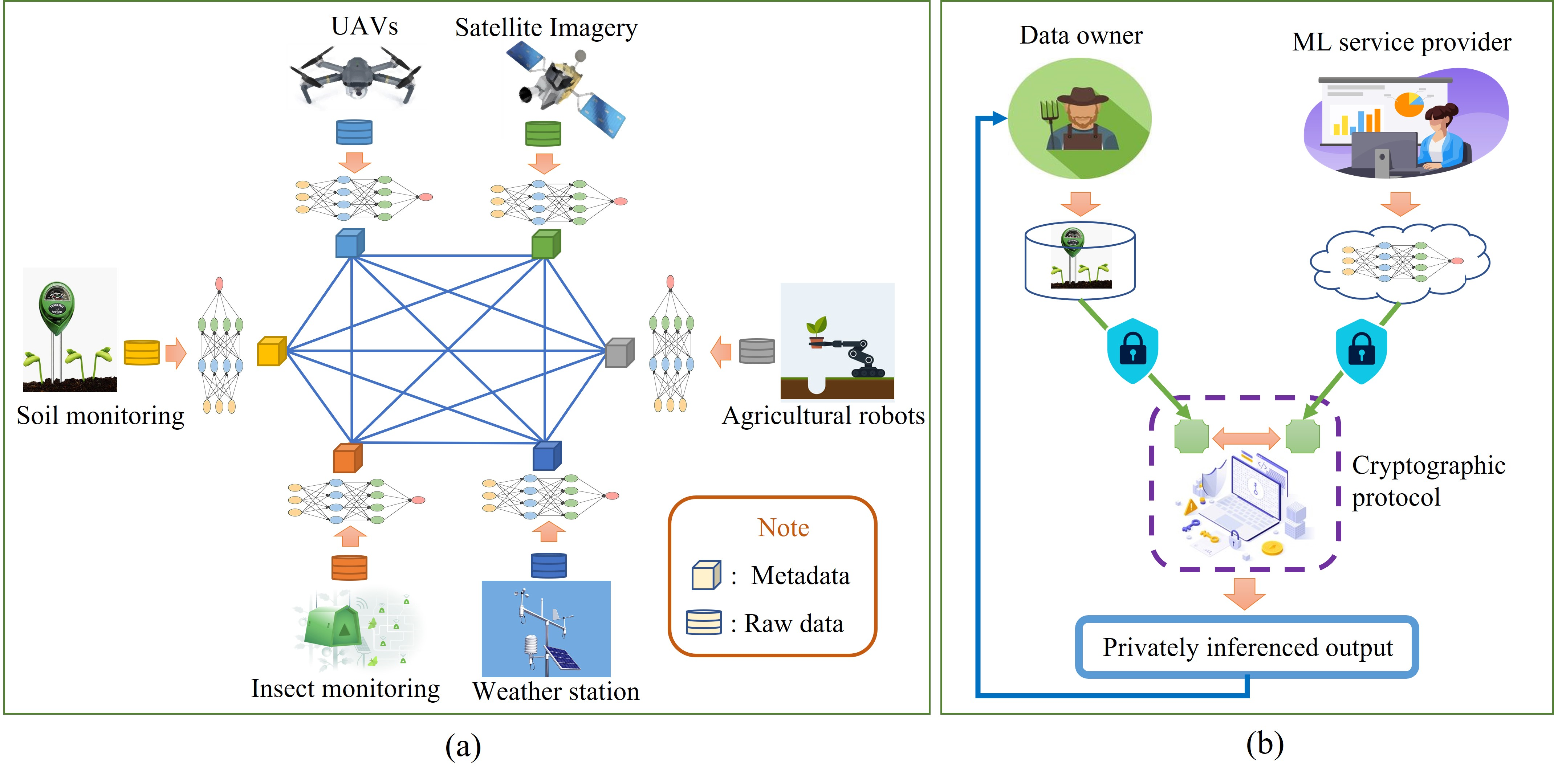}
\caption{Privacy preserving techniques while using machine learning: (a) preserving data privacy by considering peer-to-peer communication of metadata (instead of raw data) to train a machine learning model, (b) private inference by back-and-forth communicating encrypted information between data owner and ML service provider.}
\label{privacy preserving}
\end{figure}

On the other hand, in decentralized distributed learning, parties have peer-to-peer communication to exchange metadata; each party aggregates the received model parameters and updates the model. Decentralized learning architectures recently attracted attention and addressed the challenges in conventional centralized techniques, such as server latency, single point on failure, and traceability ~\citep{jiang2017collaborative, esfandiari2021cross, nadiradze2021asynchronous}. To visually present how DML can greatly reduce the privacy risk, Figure \ref{privacy preserving} (a) shows an instance of applying decentralized ML in agriculture in which each party has its private data collected from a specific IoT device and trains the model through collaborative communication of metadata based on the defined network topology (in this case a fully connected network). ~\citet{esfandiari2021distributed} used decentralized distributed learning to train an autoencoder to find anomalies in maize data. This study showed finding anomalies in private DML settings helps detect plants with irregular growth and realize probable issues during the data collection (e.g., tilted camera, rain on the lens). 

Besides DML, which naturally gives a certain level of data privacy, encryption and obfuscation are also well-known ML schemes for privacy preservation. Encryption (cryptography-based method) and obfuscation can be applied to training data and/or ML model ~\citep{gai2016privacy, juvekar2018gazelle, mishra2020delphi} (in some cases, neither the data owner nor the ML model provider does not wish to share their private parameters). In the literature, the cryptographic and obfuscation protocols include homomorphic encryption, additive secret sharing, differential privacy, and garbled circuits ~\citep{aono2017privacy, duan2022privacy, dwork2008differential, hussain2020tinygarble2}. However, all of these methods are computationally heavy, making the training/inference process several orders of magnitude slower. In agriculture, ~\citet{cho2021privacy} applied encryption by proposing novel deep neural network architectures for plant stress phenotyping that were geared towards preserving the privacy of both the user and service provider. In this method, some neural network architectures were redesigned by minimizing the number of nonlinear operations (ReLu function) to increase the inference time of encrypting model ~\citep{cho2021sphynx}. Their approach also addresses data trustworthiness, empowering farmers for SCFs. Figure \ref{privacy preserving} (b) represents the encryption scheme where privacy preserves for both the user and service provider. 

Moreover, DML and encryption techniques together were widely studied for privacy-preserving purposes  ~\citep{zhu2021distributed, mandal2019privfl, hao2019towards, zhao2019privacy, mcmahan2017communication}. In this regard, homomorphic encryption and/or differential privacy are commonly combined with FL. In the agriculture setting, ~\citet{durrant2022role} show that the privacy concern in the agri-food sector can be overcome by using federated and model-sharing machine learning as well as applying differential privacy methods. The intrusion could also be detected through an FL-based gated recurrent unit neural network algorithm (FedGRU) using the encoded data ~\citep{kumar2021pefl}. Overall, combining DML and encryption methods would be a very interesting future topic, especially in SCFs. 

\section{The Human Dimensions (Social Sciences, Economics)}
\label{S:4}
During the last two decades, on-farm research became popular due to growing and intersecting interests among farmers, agronomists, and the research community. Numerous farmer networks were organized across the country for community-based on-farm evaluations of new and existing agronomic practices~\cite{farmer_Network_Design}. Organized groups of local farmers who use GPS-enabled equipment and treatment protocols developed by researchers to conduct on-farm trials/experiments on their farms, apply treatments within their fields, and collect data~\cite{doi:https://doi.org/10.2134/precisionagbasics.2016.0096}. New analytical approaches were also developed to enhance analyses and interpretations of on-farm research data and develop dynamic decision aid tools~\cite{laurent2019framework,kyveryga2019farm}. In addition, new community-based engagement and evaluation methods were adopted by public university extension personnel and private industry agronomists~\cite{thompson2019farmers}. 

\subsection{Net-benefits of data-sharing platforms}
Increased profitability and risk reduction are some of the benefits of digital agricultural innovations to farmers. However, there are also costs associated with adopting such new technologies. These may include the cost of investing in the required technological infrastructure; the cost associated with data collection and management; the cost of analyzing the data; and the cost of sharing the data output~\cite{wysel2021data}. It is important to assess the overall cost of participating in a socio-technical platform and compare it with the potential economic gains. 

\subsection{Social barriers to data-sharing}
The adoption of technologies is also impacted by various social and ethical barriers. Despite the promise of digital agricultural technologies and data-sharing platforms, farmers are often reluctant to engage with these technologies due to issues surrounding data ownership, control, and usage~\cite{wiseman2019farmers}. Among other concerns, some farmers may be worried that technology providers could make profits off their data or share their data with third parties; and in some cases, there may be a lack of trust in the innovation itself in particular contexts of change, such as climate change~\cite{wiseman2019farmers,jakku2019if}. Also, there is a general lack of trust in data operators due to unequal power. Potential causes for farmers’ uncertainty about their rights to data ownership and usage is the lack of clarity in the terms and conditions of the agreements with service providers—not to mention the risk of potential data breaches (i.e., the risk that confidential, protected or sensitive information could be stolen or shared with third parties). Hence, there is a need for agricultural innovations that are trusted, salient and actionable for farmers to be willing to adopt them~\cite{valdivia2018new}.

\subsection{Agricultural community and the practice of farming}
The concept of the practice of farming encompasses clearly defined values, institutions, and policies~\cite{valdivia2012between}. This concept has been used to understand farmers’ behavior and actions in farming during times of policy changes to learn about how farmers transition to a new context~\cite{shucksmith1993farm,shucksmith2002future}. It has also informed the introduction of new practices, like the case of agroforestry in the context of traditional commodity farming~\cite{raedeke2003farmers,valdivia2012between}, the development of genetically modified crops~\cite{oreszczyn2010role,valdivia2014using}, and the creation of new technologies such as remote sensing with drones for agriculture~\cite{valdivia2018new}. 

Values and motivations differ in the practice of farming according to the types of crops, the nature of the market farmers engage in, and the institutions in place that support the practice~\cite{valdivia2012between}. In commodities like corn and soybeans, the field of farming has an established network of organizations and institutions that support the practice. Within it, “habitus” consists of the shared values in the practice~\cite{raedeke2003farmers, glover2010capital,valdivia2021human}. The development of new tools can benefit from the understanding of these values, the role of existing institutions that are part of the practice, and the adaptive capacity of the decision-makers. It requires an understanding of the level at which these happen and what we need to learn about the characteristics of the decision makers~\cite{ostrom2009general} to inform how new tools can provide information in changing contexts that can be trusted and acted upon~\cite{valdivia2018new, duarte2018family}. 

\subsection{Translational research and communities of practice}
According to Woolf (2008),~\cite{woolf2008meaning}, a translational research process is one that applies to advances in science toward the development of new technologies and processes, as well as guarantees that the products of research reach their intended population. By engaging the various stakeholders in participatory activities and sessions, the translational research process promotes learning, responds to challenges, and identifies opportunities. There are advantages to engaging the decision-makers from the onset of the innovation. This is critical in contexts of change. For example, simulations of pests in the context of climate change have shown that trusted and timely decisions increase the probability of success in reducing the impact of climate change. In the case of farming, the scale of the farm and that of the community and landscape as the scale of analysis, are the framework for the study of networks~\cite{garrett2011complexity,garrett2013effects}.  Participatory processes are means to facilitate learning and the development of trust~\cite{gilles2009local,patt2005effects,valdivia2009human}, as well as the co-production of knowledge~\cite{yager2019socio}.

It is argued that learning occurs as a result of social processes, and not merely through internalized cognitive processes; i.e., learning is the product of interactions between people, which lead to sharing of experiences and the creation of new knowledge~\cite{oreszczyn2010role}. Therefore, communities of practice (CoPs), defined as groups of people who  share a common activity, pursuit, or concern, are important in driving processes of change~\cite{oreszczyn2010role,valdivia2014using}. The literature suggests that farmers’ own experience with innovation, as well as the experience of their neighbors, help in decreasing the unfamiliarity with that innovation—which often acts as a significant barrier to adoption~\cite{foster1995learning}.  

In essence, the gains of learning by doing and knowledge spillovers are harnessed to a greater extent in an environment of information sharing. A CoP can help maximize farmers’ understanding of the benefits, their trust in the information source, and their eventual acceptance of the technology. Furthermore, a CoP encompasses the innovation pathway engaging all stakeholders—including feedback processes between scientists and farmers, and to other stakeholders in the practice of farming from both the public and private sectors~\cite{garrett2013effects,valdivia2014using,valdivia2018new}.

Moreover, there is a need to understand the types of incentives that farmers will respond to. There is evidence that farmers who state that privacy is important to them could still be incentivized to participate in data-sharing platforms~\cite{turland2020farmers}. As a result, it is important to investigate various factors that could potentially boost participation rates in socio-technical platforms, as well as better understand the issues that influence the adoption and overall utilization of digital agricultural technologies.

\subsection{Behavioral factors influencing agricultural technology adoption}

Traditionally, agricultural adoption research has sought to explain adoption behavior in relation to extrinsic factors such as economic, institutional, and household-specific factors~\cite{mwangi2015factors}. A more recent strand of literature has included social networks, learning, and other behavioral forces (e.g., beliefs, risk, and trust) as determinants of adoption~\cite{chavas2020uncertainty,marra2003economics}. For example, farmers belonging to a social group or organization are more likely to share information and engage in social learning about the technology~\cite{katung2008community}, hence increasing their likelihood to adopt the technology. However, social learning may also impact technology adoption negatively by increasing the likelihood of free-riding behavior among farming neighbors~\cite{bandiera2006social}. 

\begin{figure} [h]
\includegraphics[scale=.45]{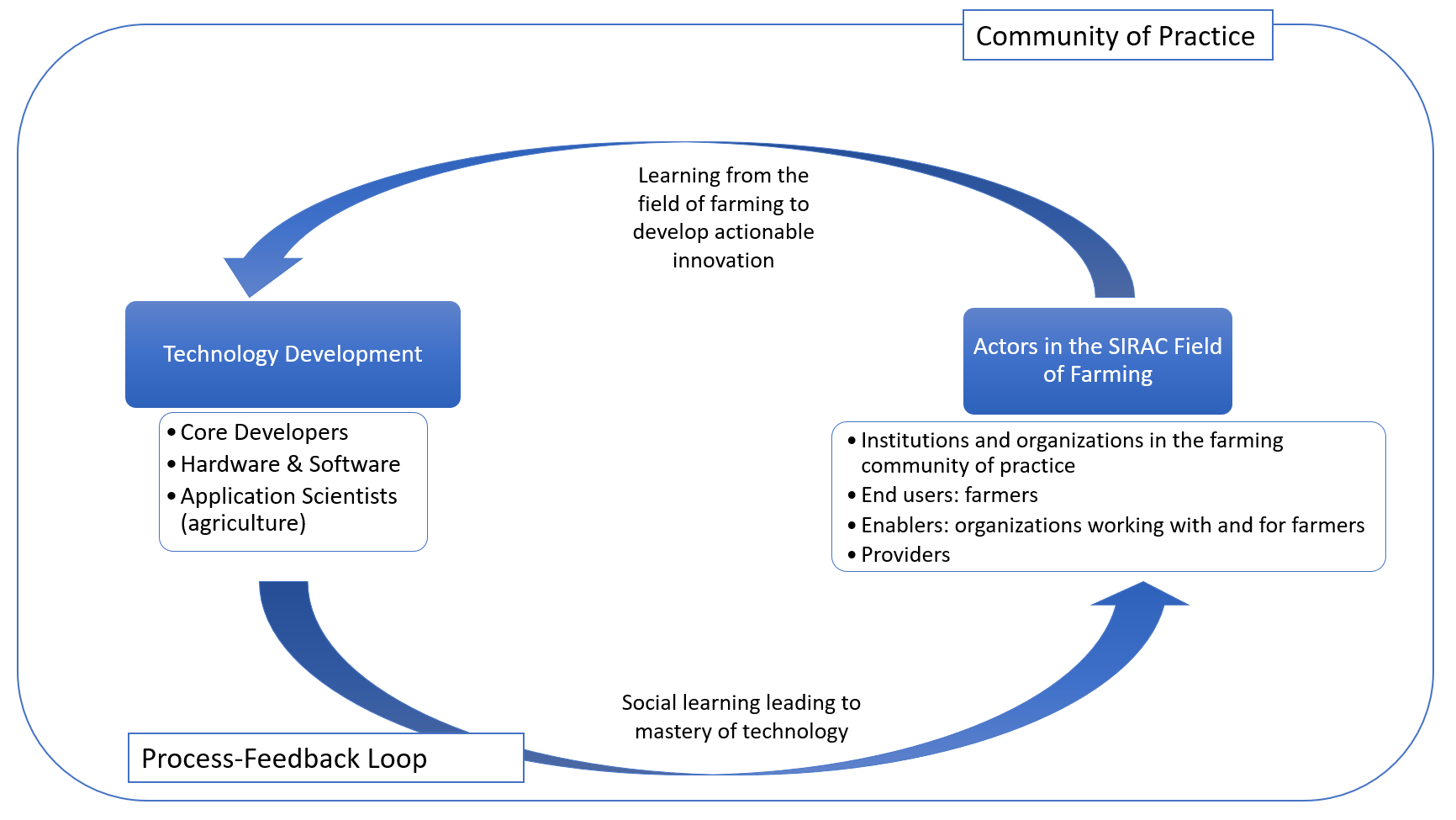}
\caption{\small{\textit{\textbf{{Translational Research Process Within The Farmer Network, such as SIRAC Community of Practice}}}}}
\label{fig:adoption_loop_high}
\end{figure}

In the context of agricultural technology adoption, different types of learning can occur simultaneously as the technology is being developed and deployed~\cite{rosenberg1982inside}. Figure~\ref{fig:adoption_loop_high} shows the technology adoption loop.  First, the developers can learn how to improve their technology through feedback from potential end-users (i.e., farmers). Second, the end-users (farmers) can improve their mastery of the adopted technology over time through personal user experience and through information acquired from various sources, including established social networks, extension specialists, and other farmers with previous experience using the technology. This creates a feedback loop process encompassing all actors in the agriculture ecosystem, from scientists and developers to potential end-users.

While social learning has received some attention in the agricultural adoption literature, the social preferences discussed extensively in other behavioral fields, including behavioral economics, have received little attention. In this regard, behavioral evidence demonstrates that other-regarding preferences, such as altruism and social norms about fairness, impact decision-making regarding technology adoption~\cite{sheeder2011empathy,chouinard2008will}. Modeling the influence of social networks on agricultural adoption decisions, beyond simple information diffusion, holds significant potential to improve the effectiveness of the deployment and adoption of agricultural innovations~\cite{streletskaya2020agricultural}. For instance, it can help technology developers understand to what extent farmers would conform to social norms as they see other farmers adopt new technology, and which are characteristics of the social networks they are more likely to join. \newline
Cooperative behavior is another factor that impacts technology adoption, especially in settings that require collective action. In this regard, the behavioral literature has studied how beliefs, trust, and risk preferences shape cooperative behavior. The motivation to study the relationship between cooperation and trust is that, for a conditional cooperator, the decision to contribute to the formation of a public good (a technological platform in our context) requires some trust in the cooperativeness of others~\cite{leonard2010social,anderson2004social,gachter2004trust}. Such trust is obviously related to the individual belief held about others’ cooperativeness. In the context of agricultural technology adoption, for a farmer who is not a free rider, contributing to the creation of the technological platform (by sharing farm data) without knowing how many other farmers are going to contribute can be viewed as a decision under risk. Thus, each farmer’s expectations about the behavior of others is a crucial determinant of contribution outcomes~\cite{cardenas2017fragility}. Farmers who are more risk averse may choose to contribute less to compensate for the risk of others, not contributing~\cite{charness2009cooperation,teyssier2012inequity,dannenberg2015provision}. For example, Liu~\cite{liu2013time} elicits risk preferences from Chinese cotton farmers and finds that risk-averse and loss-averse farmers tend to be late adopters of Bt-cotton.  Modeling cooperative behavior in the context of technology adoption is important as farmers are heterogeneous in terms of their risk aversion and trustworthiness and are more likely to join trust-supporting social networks~\cite{attanasio2012risk}. This is more so when the resolution of the product depends on how many contribute to the platform. 



\subsection{The Economics of learning with SCFs}

The foundation for the economics of SCFs are models of learning~\cite{foster1995learning,lucas1988mechanics,acemoglu2010introduction, ConleyUdry2010}. Farmers face risks and uncertainty attached to new farming practices or technologies because the suitability of the new farming methods depends on the farmer's experience, knowledge, skill, and area-specific climate and agronomic conditions. The lack of reliable and persuasive sources of information about new technologies, their expected benefit, and how to apply them efficiently impedes changes in farming behavior~\cite{Moore2008}. As a result, farmers' adoption decision requires learning, and unfortunately, learning can be costly. Acquiring, validating, analyzing, and applying new information requires time and expertise not always available to farmers. SCFs can accelerate technological adoption by reducing learning costs~\cite{foster1995learning, ConleyUdry2010,krishnan2014neighbors,carvalho2014input}. 
For example, in a recent prominent social learning analysis, economists show that network-based technology diffusion is cheaper than extension programs~\cite{benyishay2019social}. SCFs can outperform traditional agricultural extension programs if the information disseminating nodes are incentivized to transfer information about the new technology. When leading farmers were given a small incentive to distribute information to others through the network, farmer-to-farmer learning programs outperformed the government extension program in increasing farmer knowledge of new technologies. The advantage of SCFs can be significantly large if the traditional extension program requires the agents to regularly visit rural areas, especially when the extension position is remote.

\begin{figure}[h!]
\begin{subfigure}{0.4\linewidth}
  \includegraphics[keepaspectratio=true,scale=0.4]{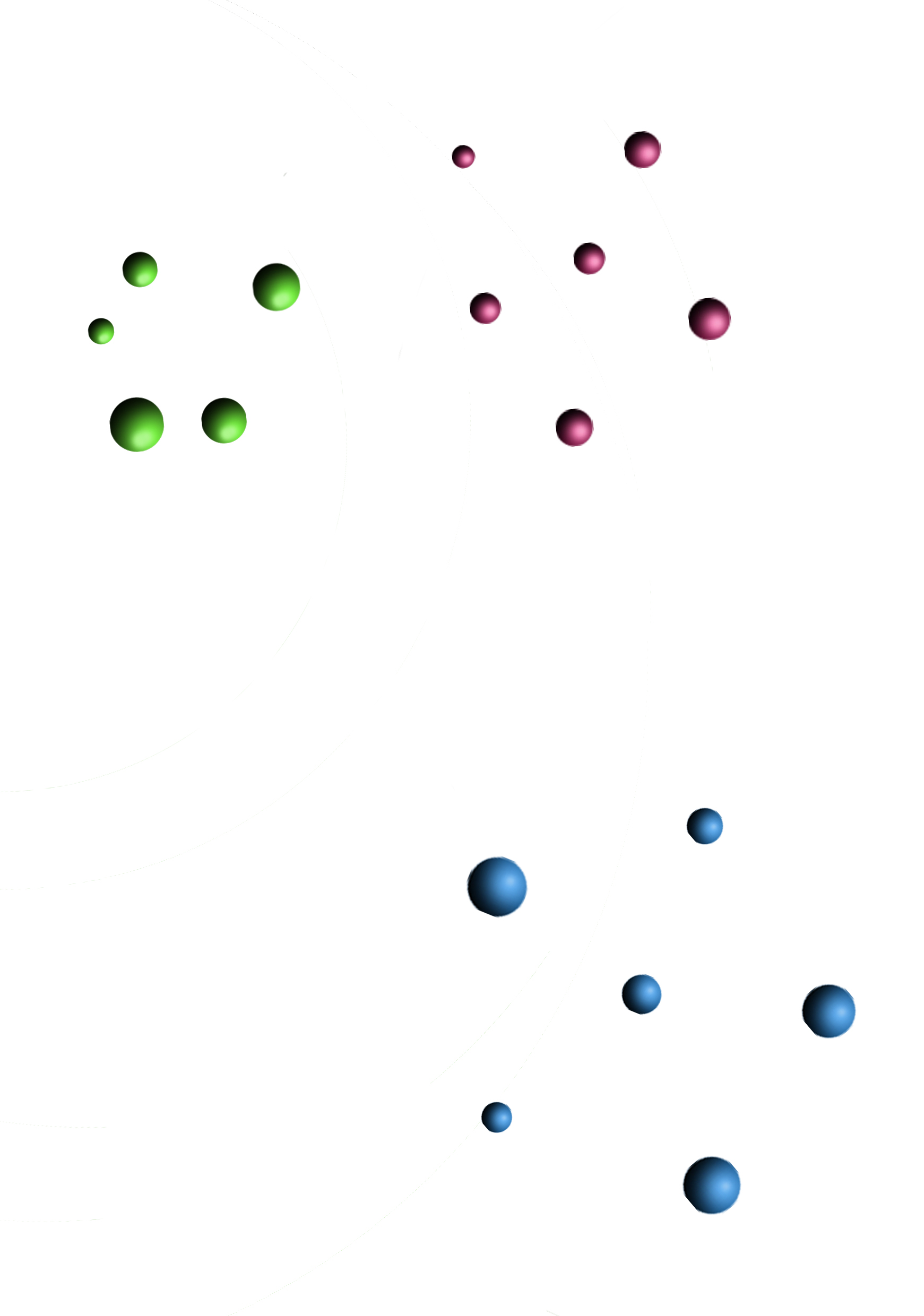}
  \caption{Traditional Scouting; No network; }
  \label{f3a}
  \end{subfigure}\hfill 
\begin{subfigure}{0.5\linewidth}
  \includegraphics[keepaspectratio=true,scale=0.4]{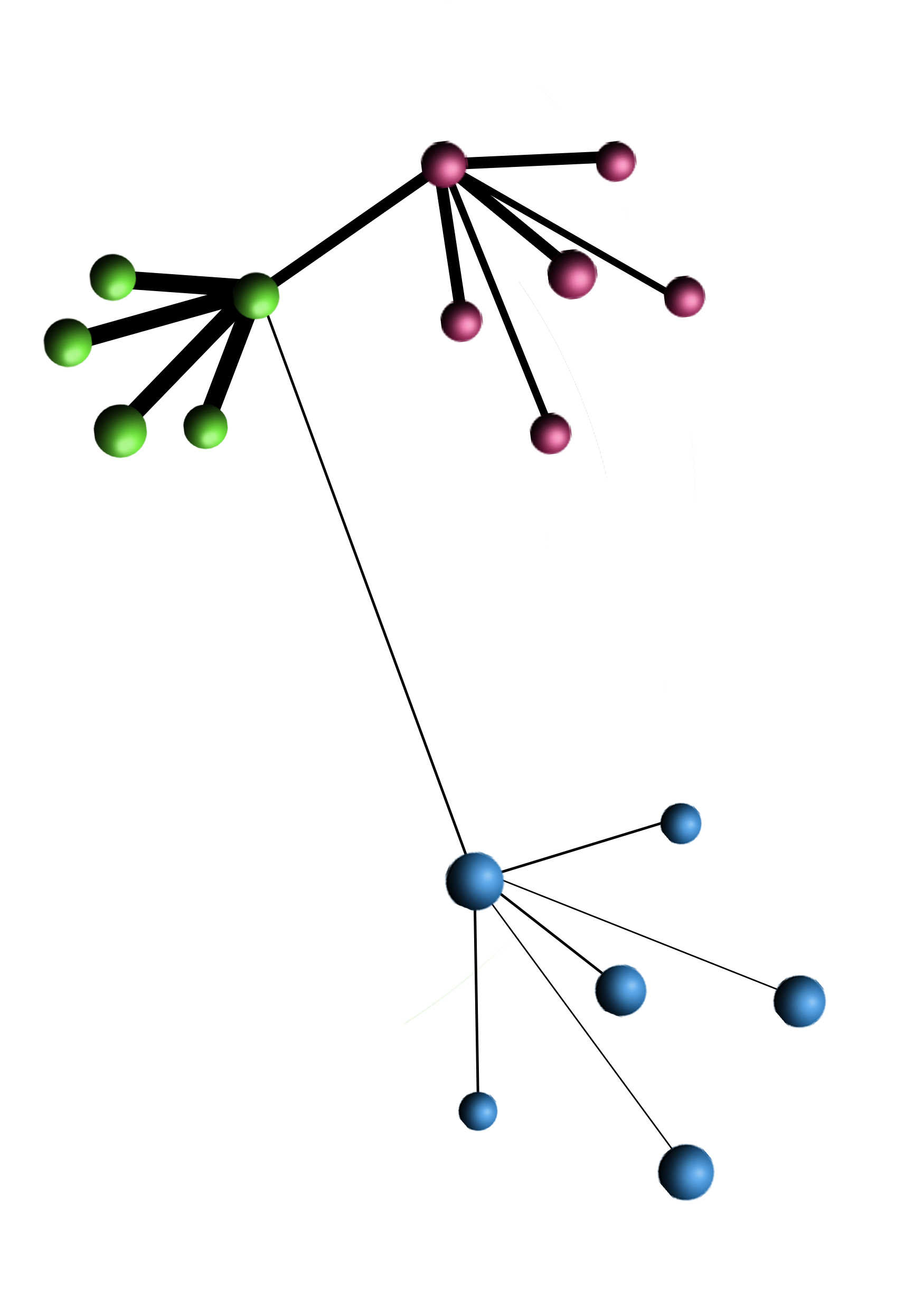}
  \caption{New pest management technology; with network.}
  \label{f3b}
\end{subfigure}
\caption{Size of the circular shapes indicates heterogeneity in agricultural production. The different colors stand for similar farms located in the same geographic cluster. In panel b, the lines indicate the connections between farmers. The lines' thickness indicates the precision of the signals.}
\label{fig:f3}
\end{figure}

Figure~\ref{fig:f3} illustrates the basic structure of a learning model for assessing the benefits of SCFs for adopting novel pest management technologies. We compare two settings with different learning mechanisms. On the left side, Figure~\ref{f3a} shows the setting for the traditional pest management technique where there is no farmer network. Pest management involves uncertainty about the amount of pesticide and the timing of the application, and farmers manage uncertainty based on their prior beliefs and their limited knowledge about the new technology. This lack of expertise leads to high uncertainty and high variability in yields and profits. For example, large uncertainty and variability about yield gains from pesticide applications have been reported~\cite{Martin2022}. The different sizes of circular shapes in the figure represent this heterogeneity in agricultural productivity.

The second setting illustrated in Figure~\ref{f3b} has a network of farmers with three clusters represented by different colors. In this setting, farmers use a new pest detection technology, digital cameras that continuously monitor selected farms for the degree of pest infestation. These digital cameras replace traditional scouting. Farmers can then learn from their peers through communication signals sent through the network. The critical parameters in this model are the precision or learning value of the signals, represented in panel b by the thickness of the connecting lines. For example, thicker lines, such as in the green cluster, mean that the signals among this homogeneous group of neighboring farmers are very informative. Using the distance between farmers to measure homogeneity and signal precision is common. For example, a receiving farmer may not trust the informational signal received from a distant farmer. The value of the SCF in the learning model illustrated in Figure~\ref{fig:f3} is measured by the reduction in uncertainty and the increase in farm profitability resulting from the higher frequency of precise network signals.

Finally, learning models can also capture the flow of unreliable information. For example, in the absence of network-based learning and expert advice, farmers may obtain information on the farming practices through word of mouth, generic broadcast programming, or agricultural input dealers, who may be poorly informed or have incentives to provide misleading information on the product, time of application, and efficiency. External agents such as pesticide retailers and commercial companies can shape the farmers' pest management decisions~\cite{Moore2008}.~\cite{Moore2008} shows that the flow of information from the retail network often plays the dominant role in pest management decisions and may negatively impact the farmers' awareness of and willingness to adjust the input use or to adopt the new farming technology. Digital monitoring and SCF networking technology have the potential to enhance the accuracy of the information.

\section{Conclusion - Transferability,  Scalability, and Adoptability of SCFs}

An SCF network has a high potential for transferability and scalability. It can be established amongst farmers who are geographically neighbors. However, the network can have the flexibility to expand to dissimilar cropping systems, management practices, soil and climate conditions, and community members. Communities can become involved through local cooperatives and/or farmer organizations within and across states. This will lead to data sharing across farms, improving their ability to engage in farm management strategies to improve productivity and broader adoption. Scalability can be facilitated with better cellular connectivity or rural broadband access to cover more significant, not necessarily contiguous, areas. The sustained usage of the SCF network will depend on buy-in from its participants and centralized community resources to support such networks' implementation and extended life cycle. There will be a need for continued education and awareness programs to ensure that the entire community is well-versed in the network's capabilities and appreciates such a network's benefits. 

There is a need for research in the area of new/improved technologies for rural connectivity and community decision-making, integrated with translational social research to address issues of adaptability, trust, and risk preferences, and economics research to justify the benefits to farmers. The research focusing on assessing social and economic incentives for farmers and other stakeholders will facilitate participation in the network, possibly through new partnerships with cooperatives, while ensuring farmer data privacy and the development of data use agreements. These SCF networks enabling rural communication technologies, including privacy-preserving distributed data analytics and machine learning tools, can apply to a broad range of cyber-physical systems applications, such as IoTs, transportation networks, and smart grids. In alignment with these technologies, the integration of Cyber-agricultural systems (CAS) offers a vision of ultra-precision agriculture, integrating improved sustainability, profitability, and technology by employing efficient sensing, AI, and robotics to address crop issues at the individual plant level~\cite{sarkar2023cyber}. These research topics will require a trans-disciplinary team of researchers from multiple domains, including agronomy, computer science, plant pathology, behavioral science, economics, sensing and machine learning, and precision agriculture, along with integral participation of farmers to define and conduct these research, so the benefits of SCF network is immediate and useful. The usefulness of citizen science data sets~\cite{iNaturalist} for deep learning powered real-time identification of insects is an example of bridging the gap in the development of SCF~\cite{chiranjeevi2023deep, saadati2023out}. 

\section*{Author Contributions}
\textbf{}All student and post-doctoral fellow members of this interdisciplinary team worked with their faculty mentors. Each author contributed to the planning and writing of this review article.

\section*{Acknowledgements}
We thank farmer partners in the community of practice, awareness community, as well as Iowa Soybean Association.

\section*{Funding}
\textbf{}This work was supported by Smart Integrated Farm Network for Rural Agricultural Communities (SIRAC) (NSF S\&CC 1952045), AI Institute for Resilient Agriculture (USDA-NIFA 2021-67021-35329), COALESCE: COntext Aware LEarning for Sustainable CybEr-Agricultural Systems (NSF CPS Frontier 1954556), FACT: A Scalable Cyber Ecosystem for Acquisition, Curation, and Analysis of Multispectral UAV Image Data (USDA-NIFA 2019-67021-29938), and USDA CRIS Project IOW04714. 

\section*{Declaration of conflicting interests}
\textbf{Competing Interests:} None

\clearpage

\bibliographystyle{unsrtnat}
\bibliography{references}  

\end{document}